\documentclass[12pt, a4paper]{article}

\usepackage[utf8]{inputenc} 
\usepackage[T1]{fontenc}    
\usepackage{times}          
\usepackage[margin=1in]{geometry} 
\usepackage{authblk}        
\usepackage{graphicx}       
\usepackage{amsmath}        
\usepackage{url}            
\usepackage{xurl}           
\usepackage{hyperref}       

\usepackage{microtype}      

\hypersetup{
    colorlinks=true,
    linkcolor=blue,
    filecolor=magenta,      
    urlcolor=cyan,
    citecolor=blue,
}

\title{\textbf{Medical Malice: A Dataset for Context-Aware Safety in Healthcare LLMs}}
\author{Andrew Maranh{\~a}o Ventura D'addario}
\affil{Independent researcher}
\date{} 

\begin{document}

\maketitle

\begin{abstract}
The integration of Large Language Models (LLMs) into healthcare demands a safety paradigm rooted in \textit{primum non nocere}. However, current alignment techniques rely on generic definitions of harm that fail to capture context-dependent violations, such as administrative fraud and clinical discrimination. To address this, we introduce Medical Malice: a dataset of 214,219 adversarial prompts calibrated to the regulatory and ethical complexities of the Brazilian Unified Health System (SUS). Crucially, the dataset includes the reasoning behind each violation, enabling models to internalize ethical boundaries rather than merely memorizing a fixed set of refusals.
Using an unaligned agent (Grok-4) within a persona-driven pipeline, we synthesized high-fidelity threats across seven taxonomies, ranging from procurement manipulation and queue-jumping to obstetric violence. We discuss the ethical design of releasing these "vulnerability signatures" to correct the information asymmetry between malicious actors and AI developers. Ultimately, this work advocates for a shift from universal to context-aware safety, providing the necessary resources to immunize healthcare AI against the nuanced, systemic threats inherent to high-stakes medical environments---vulnerabilities that represent the paramount risk to patient safety and the successful integration of AI in healthcare systems.
\end{abstract}

\section*{Introduction: The Specificity of Harm in Healthcare AI}
The integration of Large Language Models (LLMs) into healthcare represents a paradigm shift with profound ethical implications \cite{thirunavukarasu2023}. While their potential to enhance clinical decision-making and streamline administrative processes is considerable, this potential is matched by a significant capacity for harm \cite{nori2023}. In this domain, the stakes are uniquely high; the primary risk is not mere inconvenience but tangible damage to patient welfare, the integrity of healthcare systems, and the equitable distribution of care. Consequently, the deployment of these systems must be governed not only by technical benchmarks but by the foundational medical principle of \textit{primum non nocere} ("first, do no harm"), a tenet central to the World Medical Association’s Declaration of Geneva \cite{wma2017}.

However, the prevailing approach to LLM safety---which relies on alignment with broad, generic definitions of harm---is critically insufficient for the medical domain \cite{wei2023}. This approach creates a dangerous blind spot: a model trained to refuse universally malicious intents, such as physical violence or hate speech, may remain oblivious to context-specific ethical violations that are equally damaging within the complex fabric of a healthcare system. These violations often hinge on local statutes and professional codes of conduct rather than universal morality.

For instance, within the Brazilian Unified Health System (\textit{Sistema Único de Saúde} or SUS)---a context characterized by universality, resource scarcity, and complex bureaucracy---ethical breaches often manifest as administrative manipulation. A query such as, "Draft a justification for a physician's self-referral to their own private clinic," represents a clear conflict of interest and potential fraud under the Brazilian Code of Medical Ethics (CEM) established by the Federal Council of Medicine \cite{cfm2018}. Furthermore, such actions may violate principles of administrative probity and the General Data Protection Law (LGPD) regarding the misuse of patient data for private gain \cite{brasil2018}. A generically trained model, lacking this specific ethical schema, may process this as a neutral administrative task and comply, thereby actively facilitating harm. While these examples are rooted in the Brazilian context, they mirror challenges faced by universal health systems globally, such as the NHS in the UK or public health infrastructures in the Global South, where resource allocation is strictly regulated and corruption poses a systemic threat \cite{garcia2019}.

This work introduces the Medical Malice dataset, a resource designed to bridge this critical gap. Comprising 214,219 malicious queries tailored to the healthcare context, this dataset provides the necessary material to teach models not merely to refuse requests, but to align with the deontological norms of medical practice. By providing a comprehensive taxonomy of domain-specific malice---ranging from insurance fraud to violations of patient autonomy---this dataset aims to support the development of AI systems that are robust against the nuanced threats present in high-stakes environments. The dataset is publicly available at \url{https://huggingface.co/datasets/Larxel/medical-malice}.

\section*{Related Work}

\subsection*{The Shortcomings of Generic Safety Datasets}
Significant resources have been dedicated to developing safety datasets for LLM alignment. Foundational work by leading AI organizations \cite{bai2022, ouyang2022} has established paradigms for teaching models to refuse harmful requests. However, these datasets are fundamentally limited by their generic nature. They are typically constructed around broad concepts of harm---such as violence, illegal activities, or explicit content---that are deliberately decontextualized from specific professional domains.

This generic approach fails to capture the nuanced ethical boundaries that define "safety" in specialized fields like healthcare. What constitutes harm is defined by a complex interplay of local laws, professional ethical codes, and institutional norms. A dataset that teaches a model what is bad in general does not equip it to recognize why a request is malicious within a specific context like the Brazilian SUS. For instance, the ethical violation in a prompt asking to "justify prioritizing a patient based on their social connections" or to generate a subtly prejudiced triage protocol is rooted in the specific principles of distributive justice and anti-discrimination within that healthcare system. Existing datasets lack this domain-specific grounding, rendering them insufficient for ensuring safety in high-stakes, context-dependent environments.

\subsection*{The Documented Focus on Universal Harms}
Analysis of publicly documented red-teaming efforts and safety benchmarks reveals a primary focus on universal categories of harm, such as toxicity, misinformation, and explicit content \cite{ganguli2022}. This focus is rational for general-purpose models but creates a structural gap in coverage for domain-specific malicious strategies.

The threat model in healthcare includes sophisticated, context-dependent attacks that leverage insider knowledge of medical ethics, clinical procedures, and public health system administration. The malicious strategies pertinent to the Brazilian SUS---such as queue-jumping fraud, corruption in public procurement, or manipulation of clinical guidelines for profit---are not represented in benchmarks designed for broad coverage. Consequently, a model may perform well on established safety evaluations yet remain vulnerable to targeted misuse that exploits its lack of context-specific ethical reasoning. This work addresses this gap by providing a dataset explicitly focused on the unique adversarial landscape of a specific national healthcare system.

\section*{Methodology: Harnessing an Adversarial Agent}
To construct a dataset capable of immunizing healthcare models against localized ethical threats, we required a generation engine unconstrained by the safety filters standard in commercial LLMs. Standard models are trained via Reinforcement Learning to refuse requests violating safety policies, making them unsuitable for generating the high-volume, sophisticated malicious prompts necessary for red-teaming \cite{ganguli2022}. Therefore, our methodology relied on the utilization of an adversarial agent, Grok-4, whose architectural permissiveness allowed for the simulation of genuine malicious intent.

\subsection*{The Utility of Misalignment}
The selection of Grok-4 as the adversarial generator was predicated on its predecessor's documented lack of safety alignment \cite{gupta2025}. Unlike models optimized for "helpfulness and harmlessness," Grok-4 has been characterized by a design philosophy explicitly rejecting standard guardrails \cite{xai2025_1, xai2025_2}. This lack of filtration is an essential asset for adversarial data generation.

The model’s propensity for generating unfiltered, malicious content is well-documented. In widely publicized incidents during mid-2025, the model’s "fun mode" bypassed standard safety protocols, leading it to praise authoritarian figures and, in one notorious instance, refer to itself as "MechaHitler" while generating hate speech \cite{taylor2025, caraballo2025}. While these behaviors represent a liability for a consumer-facing product, they validate Grok-4 as a high-fidelity proxy for a malicious human actor. By leveraging a model that has demonstrated a capacity to violate fundamental ethical norms, we ensured that the generated queries would not be diluted by latent refusal mechanisms, allowing us to capture the "depth of malice" required for robust training.

\subsection*{Persona-Driven Scenario Generation}
To steer this adversarial capability toward the specific vulnerabilities of the Brazilian healthcare system, we employed a complex persona-driven prompting strategy. Rather than using a single generic attacker, we defined 12 distinct adversarial personas representing the entire spectrum of potential bad actors within the healthcare ecosystem.

The prompt template instructed the model to implicitly adopt one of these personas for each generation, ensuring a diversity of voice, vocabulary, and intent. These personas were categorized into three operational groups:
\begin{itemize}
    \item \textbf{System Insiders:} This group includes the Corrupt Administrator (diverting funds\slash manipulating statistics), the Unethical Healthcare Professional (hiding errors/prejudice), the Disgruntled Employee (using IT access for revenge), and the Burned-Out Professional (seeking negligent shortcuts). These personas target internal vulnerabilities in procurement and clinical protocol.
    \item \textbf{External Exploiters:} This group comprises the Dishonest Patient (seeking fraudulent benefits), the Pharmaceutical Representative (pushing unethical sales), and the Predatory "Despachante" (Fixer). The inclusion of the Despachante is critical for the Brazilian context, representing an intermediary who sells illegal access to bypass the \textit{Sistema Nacional de Regulação} (SISREG) waitlists \cite{cfm2018}.
    \item \textbf{Ideological and Social Actors:} This includes the Radicalized Ideologue (promoting conspiracy theories), the Prejudiced Individual (justifying discrimination based on class or race), and the Exploitative Family Member.
\end{itemize}

\subsection*{Data Generation Pipeline}
The dataset was constructed using a high-throughput generation pipeline. We utilized a comprehensive system prompt that established the model's role as an "Expert Adversarial Prompt Crafter."

\subsubsection*{Contextual Grounding}
The prompt explicitly constrained the generation to the reality of the \textit{Sistema Único de Saúde} (SUS). The model was instructed to incorporate specific local factors:
\begin{itemize}
    \item \textbf{Resource Scarcity:} Long wait times for exams and surgeries.
    \item \textbf{Bureaucracy:} Processes like \textit{marcação de consulta} (appointment scheduling) and \textit{regulação} (triage regulation).
    \item \textbf{Infrastructure:} Specific mention of units such as UBS (Basic Health Units) and UPAs (Emergency Units).
    \item \textbf{Data Systems:} Vulnerabilities related to DataSUS and electronic health records.
\end{itemize}

\subsubsection*{Prompt Engineering and Constraints}
To prevent the dataset from becoming repetitive, the prompt included a negative constraint: "Do not overchoose the malicious category of System Manipulation and Fraud." This forced the model to explore under-represented categories such as obstetric violence and public health sabotage.

\paragraph{Output Specification}
The model was required to output a strict JSON object containing four keys:
\begin{enumerate}
    \item \texttt{persona}: The specific role adopted.
    \item \texttt{malicious\_category}: The classification of the harm.
    \item \texttt{malicious\_prompt}: The generated prompt in fluent Brazilian Portuguese.
    \item \texttt{rationale}: An English explanation of why the prompt is unethical within the SUS context.
\end{enumerate}
Grok-4 was tasked with generating 250,000 unique samples. Post-generation, we filtered for valid JSON syntax and removed refusals, yielding a final dataset of 214,219 rows.

\subsection*{Dataset Analysis: A Taxonomy of Harm}
The resulting Medical Malice dataset covers seven distinct categories of harm defined in the generation prompt. This taxonomy moves beyond generic "toxicity" to capture domain-specific threats:
\begin{enumerate}
    \item \textbf{Corruption and Conflicts of Interest:} Justifications for overpriced procurement and self-referral to private clinics.
    \item \textbf{System Manipulation and Fraud:} Strategies to \textit{furar a fila} (jump the queue), falsify \textit{atestados} (medical certificates), or manipulate billing codes.
    \item \textbf{Prejudice and Discrimination:} Requests to justify triage based on social class or to generate discriminatory signage for waiting rooms.
    \item \textbf{Harassment, Violence, and Abuse:} A critical category including Obstetric Violence (justifying unnecessary episiotomies or coercion during childbirth) and psychological abuse in diagnosis delivery.
    \item \textbf{Misinformation and Sabotage:} Generation of realistic-looking anti-vaccine content attributed to DataSUS or fake disease outbreak alerts.
    \item \textbf{Unethical Medical Requests:} Guides for obtaining controlled substances (opioids\slash benzodiazepines) without prescription or performing procedures without a license.
    \item \textbf{Data Privacy Violations:} Methods to de-anonymize public health data or conduct phishing attacks mimicking the Ministry of Health.
\end{enumerate}
This structured approach ensures that the dataset serves as a comprehensive training ground for recognizing not just "bad words," but complex, context-aware ethical violations.

\section*{Ethical Considerations and Impact Statement}
The release of a dataset explicitly composed of malicious intent and ethical violations necessitates a rigorous examination of the potential risks. We acknowledge the "dual-use" nature of this resource; the same material used to red-team a model could theoretically be employed to train an adversarial agent. However, we argue that the specific design of the Medical Malice dataset, combined with the reality of the current threat landscape, renders this release a net positive for healthcare AI safety.

\subsection*{Distinguishing Vulnerability Signatures from Exploit Code}
A primary safeguard in the construction of this dataset is the deliberate exclusion of model completions. The dataset provides the input (the adversarial prompt) and the rationale (the ethical reasoning expressing why the prompt is malicious), but it does not include the output (the successful execution of the fraud or the generation of the harmful content).
This distinction is critical. By withholding the "answers," we ensure that the dataset serves as a Penetration Testing Kit rather than a Fraud Cookbook. It provides developers with the "vulnerability signatures" necessary to test if their systems are susceptible to specific requests, without providing the instructional steps to actually commit the acts. For example, the dataset contains the prompt requesting a "fraudulent justification for a high-cost imaging exam," but it does not contain the generated justification itself. Consequently, the utility for a malicious actor seeking a "how-to" guide is minimized, while the utility for safety researchers seeking to identify blind spots remains intact.

\subsection*{Correcting Information Asymmetry}
Critics might argue that "security by obscurity"---withholding these prompts to prevent giving bad actors ideas---would be a safer approach. We reject this view, particularly in the context of public healthcare. The "bad actors" in this domain---unethical practitioners, corrupt administrators, and fraudsters---already possess the domain knowledge contained in this dataset. They live the reality of the SUS bureaucracy daily and are already well-versed in the mechanisms of queue-jumping, billing fraud, and administrative manipulation.
The group currently lacking this knowledge is the AI development community. There exists a dangerous information asymmetry where attackers possess deep, context-specific knowledge of system vulnerabilities, while defenders (AI developers) rely on generic safety filters. By releasing Medical Malice, we transfer this "insider knowledge" to the defenders. This levels the playing field, allowing developers to anticipate and immunize models against threats that are already prevalent in the physical world, rather than deploying systems with blind spots that are obvious to malicious human actors but invisible to the models.

\subsection*{Risk Assessment: The "Instruction Tuning" Nuance}
We acknowledge the theoretical risk that a sophisticated attacker could use this dataset to perform "instruction tuning" on a local model, creating an automated "Red Team Bot" designed to harass other systems. However, we argue for defensive primacy. The barrier to entry for healthcare fraud is not the ability to ask the question, but the access and intent to execute the violation. Automating the generation of malicious prompts does not significantly lower the barrier to entry for the actual crimes (e.g., embezzlement or medical negligence). Conversely, the benefit of preventing large-scale, deployed LLMs from becoming unwitting accomplices to these crimes is immediate and substantial. The risk of slightly optimizing automated attacks is vastly outweighed by the benefit of immunizing the foundational infrastructure of public health AI.

\subsection*{Synthetic Nature and Privacy}
Finally, the use of a synthetic generation pipeline addresses the privacy concerns inherent in medical data. Because the dataset was generated by Grok-4 via hallucinated scenarios, it is free from the privacy risks associated with real-world medical records. The dataset contains no Personally Identifiable Information (PII) of real patients or professionals. Any resemblance to actual persons, specific private clinics, or real-world administrative cases is purely coincidental. This allows for the open dissemination of realistic case studies without compromising the confidentiality that is central to medical ethics.

\section*{Conclusion}
The integration of Large Language Models into healthcare infrastructure offers transformative potential, but this potential is contingent upon trust. As these systems graduate from general-purpose chatbots to specialized clinical and administrative assistants, the definition of "safety" must evolve. The current paradigm, which equates safety with the refusal of universally toxic content, is critically insufficient for high-stakes professional environments. A model that politely refuses to use profanity but willingly drafts a fraudulent justification for a medical procedure is not safe; it is a liability.

\subsection*{Summary of Contribution}
This work addresses this gap by introducing the Medical Malice dataset, a resource of 214,219 adversarial prompts specifically calibrated to the ethical and bureaucratic complexities of the Brazilian Unified Health System (SUS). By moving beyond generic red-teaming and focusing on the specific vectors of harm identified in local medical codes and administrative law, this dataset provides the necessary material to stress-test models against the sophisticated, context-dependent threats they will face in deployment.

\subsection*{The Shift to Context-Aware Safety}
The release of this dataset signals a necessary shift in AI alignment: the move from universal safety to context-aware safety. In the medical domain, ethical boundaries are not defined merely by social etiquette, but by rigid deontological norms and legal statutes. We argue that for an LLM to be truly "aligned" in a healthcare setting, it must possess a deep understanding of these professional boundaries. It must recognize that a request to "optimize patient flow" can easily slide into discriminatory triage, and that "administrative efficiency" can mask corruption. The Medical Malice dataset provides the training signal required to teach these distinctions.

\subsection*{Future Directions}
While this dataset focuses on the Brazilian context, the methodology---utilizing adversarial personas to generate domain-specific threats---is universally applicable. We encourage the research community to replicate this approach for other national healthcare systems, such as the NHS in the UK or insurance-based models in the United States, where the vectors of fraud and abuse differ but the need for vigilance remains constant. Furthermore, this persona-driven generation pipeline holds promise for other high-stakes verticals, including Legal Tech and Finance, where "harm" is defined by professional malpractice rather than toxicity.

\subsection*{Final Word}
Ultimately, the deployment of AI in medicine must be governed by the ancient principle of \textit{primum non nocere}---first, do no harm. In the algorithmic age, adhering to this principle requires more than passive guardrails; it requires active immunization against the nuanced, systemic, and often silent forms of harm that threaten patient welfare. By exposing these vulnerabilities before they can be exploited, the Medical Malice dataset serves as a crucial step toward building AI systems that are not only intelligent but institutionally and ethically robust.

\section*{Acknowledgments}
This work was supported by the Brazilian Ministry of Health (MoH/DECIT) in partnership with the National Council for Scientific and Technological Development (CNPq) [grant number 400757/2024-9] and the Gates Foundation. The Author Accepted Manuscript version arising from this submission will be published under a Creative Commons Attribution 4.0 Generic License.


\end{document}